# Rate dependence of damage formation in metallic-intermetallic Mg-Al-Ca composites

Setareh Medghalchi[1,*], Muhammad Zubair[1,2,*], Ehsan Karimi[1], Stefanie Sandlöbes-Haut[1], Ulrich Kerzel[3], Sandra Korte-Kerzel[1]

[1] Institute for Physical Metallurgy and Materials Physics, Kopernikusstr. 14, RWTH Aachen University, 52074 Aachen, Germany.

[2] Department of Metallurgical & Materials Engineering, Faculty of Chemical, Metallurgical & Polymer Engineering, University of Engineering & Technology (UET) Lahore, Pakistan.

[3] Data Science and Artificial Intelligence in Materials and Geoscience, Fakultät für Georessourcen und Materialtechnik, RWTH Aachen University, Aachen, Germany.

*Corresponding authors

## Abstract

We study a cast Mg-4.65Al-2.82Ca alloy with a microstructure containing α-Mg matrix reinforced with a C36 Laves phase skeleton. Such ternary alloys are targeted for elevated temperature applications in automotive engines since they possess excellent creep properties. However, in application, the alloy may be subjected to a wide range of strain rates and in material development, accelerated testing is often of essence. It is therefore crucial to understand the effect of such rate variations. Here, we focus on their impact on damage formation. Due to the locally highly variable skeleton forming the reinforcement in this alloy, we employ an analysis based on high resolution panoramic imaging by scanning electron microscopy coupled with automated damage analysis by deep learning-based object detection and classification convolutional neural network algorithm (YOLOV5). We find that with decreasing strain rate the dominant damage mechanism for a given strain level changes: at a strain rate of $5 \cdot 10^{-4}$/s the evolution of microcracks in the C36 Laves phase governs damage formation. However, when the strain rate is decreased to $5 \cdot 10^{-6}$/s, interface decohesion at the α-Mg/Laves phase interfaces becomes equally important. We also observe a change in crack orientation indicating an increasing influence of plastic co-deformation of the α-Mg matrix and Laves phase. We attribute this transition in leading damage mechanism to thermally activated processes at the interface.

# 1 Introduction

Many high-performance alloys consist of two or more phases to combine and improve their respective properties in a microstructural composite. In lightweight magnesium alloys, cast microstructure with an intermetallic skeleton have been shown to lead to superior creep resistance, particularly at elevated temperature [1-3]. On the other hand, the presence of intermetallics naturally leads to decreased tensile elongation as the present $Mg_{17}Al_{12}$ and $Ca(Mg,Al)_2$ Laves phases are hard and brittle, particularly below approximately two thirds of their melting temperature [4-11]. As both low and elevated temperature regimes, that is room temperature and temperature around and above 150 °C, are of interest in application, the underlying deformation and damage mechanisms and their thermal activation need to be understood. In particular, the competition between brittle fracture and plasticity, enabled by dislocation glide or diffusional processes, may be expected to govern any macroscopic transitions in behaviour with temperature. In this work, we address this by investigating the signatures of the dominant co-deformation mechanisms as a function of strain and rate in a Mg-$Ca(Mg,Al)_2$ metallic-intermetallic alloy by means of micromechanical testing and damage classification using artificial intelligence.

Laves phases with the general formula $Ca(Mg,Al)_2$ precipitate in the Mg-Al-Ca ternary alloy system once the Al and Ca alloying content exceed the solubility limit [12]. Laves phases have an $AB_2$ stoichiometry and are known for their high temperature strength, which is always accompanied with extreme brittleness at low temperatures [13-15]. However, the Laves phases usually show plasticity at high temperatures (above the ductile-brittle transition temperature) or at small scales [5, 6, 9, 13, 15-18]. In conventional, cast Mg-Al-Ca alloys, they are usually present as thin interconnected struts with a thickness of the order of one to a few micrometre [19-21]. Their presence as interconnected reinforcement imparts good creep properties [1-3, 19, 22, 23], thermal stability [24], and adequate strength [25] to Mg-Al-Ca alloys. These alloys are thus intended for elevated temperature structural applications like automotive powertrains [26].

In Mg-Al-Ca alloys, the Mg and Laves phases have significantly different mechanical properties, which results in heterogenous deformation [8, 27, 28]. Under tensile loading or during indentation, cracks form in the Laves phase or at α-Mg/Laves phase interfaces [3, 8, 20, 27, 29, 30]. In addition to cracking, plastic deformation has also been observed in the Laves phases surrounded by α-Mg matrix [8, 16, 28, 31]. The co-deformation behaviour of α-Mg and Laves phases can be substantially affected by the strain rate, in addition to other factors like orientation relationships and temperature of deformation [28]. Strain rate and temperature are known to considerably affect the deformation and fracture behaviour of materials [32]. The flow stress or hardness decreases with decreasing strain rate and vice versa owing to thermally activated plastic deformation mechanisms and changing balance in their competition with fracture as temperature increases.

In this work, we investigate the effect of strain rate on the elevated temperature (≈170 °C) deformation behaviour of an Mg-4.65Al-2.82Ca alloy. As a consequence of deformation at different strain rates, any changes in prevalent deformation mechanism will lead to different amounts and characteristics of damage sites that are introduced in the microstructure during deformation. The Mg-Al-Ca metal-intermetallic composites possess several critical microstructural length scales from the thickness of the intermetallic struts (~1 μm) over the strut spacing in the skeleton (10s of μm) to the grain size (100s of μm). These naturally expand into all three dimensions in this cast alloy, but three-dimensional characterisation is normally limited in the sense that interrogated volume and voxel size and therefore resolution scale. We therefore choose cross-sectional analysis after deformation by scanning electron microcopy (SEM), which is able to cover a large area of the order of 1 mm² at sufficiently high resolution to encompass all important length scales for this alloy. By using deep learning to assist in the analysis of the many damage sites induced across a large cross-sectional area after deformation, we consider the formed damage statistically in terms of its major characteristics or classes of damage observed.

Deep learning methods based on convolutional neural networks serve as a tool to learn spatial hierarchies of information about the content of image automatically and adoptively through backpropagation [33]. Here, we use two image analysis tasks: feature detection, which is the task of localising all instances of a specific class in an image [34], and classification of the objects in an image [35], which is the task of assigning features of an image to specific classes [36]. In the case of this work, damage sites in panoramic micrographs are detected and classified with the objective of damage analysis. In this context, deformation induced damage sites can be detected and subsequently classified with respect to their appearance, as shown by Kusche et. al and Medghalchi et. al. for damage sites in dual phase steel [37, 38].

In this work, we explore the prevalent mechanisms of co-deformation and their dependence on strain and rate in a Mg-Ca(Mg,Al)$_2$ metallic-intermetallic composite microstructure. To this end, we use micromechanical testing and scanning electron microscopy coupled with automated image analysis and damage classification to identify and quantify the dominant damage mechanisms of brittle failure in the intermetallic and interfacial decohesion at the internal interfaces. These insights are essential to guide future material design strategies dedicated to achieving a damage tolerant microstructure with tailored strength, creep resistance and elongation to failure for a given application.

# 2 Experimental Methods

## 2.1 Sample synthesis, deformation, and imaging

A protective atmosphere of argon was used to melt the raw materials in a steel crucible using a vacuum induction melting system. Mg-4.65Al-2.82Ca (wt.% alloy) was then solidified in a copper mould as described in [39].

Dog bone shaped specimens with a 10 mm gauge length for tensile deformation were cut by spark erosion. The grinding and polishing procedure for the samples was the same as discussed comprehensively in [12]. Samples were deformed to 3, 5 and 7% global strain at a temperature of 170 °C and strain rate of 5 x $10^{-4}$/s. One sample was also deformed at 5 x $10^{-6}$/s to the intermediate strain of 5% to study the effect of strain rate on damage formation.

Post deformation microscopic analysis was done using secondary electron (SE), back-scattered electron (BSE) panoramic imaging, and electron back-scatter diffraction (EBSD) in scanning electron microscopes (SEM, FEI Helios 600i and Zeiss LEO1530). Panoramic imaging was done to cover maximum microstructural region without losing resolution or image quality. For example, the microstructural region presented in Figure 9 (a) is comprised of 255 individual images, each with 50 µm (1024 px) of horizontal field width. An acceleration voltage ranging from 8-10 kV was used for this purpose. Images were stitched using image composite editor [40]. EBSD was performed at an accelerating voltage of 20kV.

## 2.2 Automatic damage analysis method

### 2.2.1 Object detection and classification model

Yolo (You Only Look Once) as one of the popular object detection and classification algorithms has gained lots of attention in a wide range of applications such as object detection and image segmentation [41]. In contrast to two-stage detector methods, in which region proposal and classification tasks are sequentially performed, Yolov5 is a one-stage detector methods, in which region proposal and classification are solved simultaneously [42]. This algorithm is relatively easier to implement and can be trained on the entire image without the need to divide the image or filter the objects of interest [43]. Its fifth version (YoloV5), which is the most recent one at the time of doing this research, has overperformed other variants such as fast R-CNN in terms of speed and accuracy. Here, we use YoloV5s, which has one of the fastest learning rates among all other options with 37.2 validation mAP (mean average precision) @ [0.5:0.95] on COCO API dataset [41, 44] [45].

We use Yolov5s to detect and classify the damage sites in the microstructure of the Mg-Al-Ca alloy across panoramic SEM images. For further damage analysis, we used the Hough transformation to extract the inclination of cracks in the Laves phase.

### 2.2.2 Definition of damage classes

The deformed microstructure is composed of Mg matrix with the Ca(Mg,Al)$_2$ Laves phase, and also contains deformation features such as slip lines, twins and cracks or interfacial decohesion sites. The cracks evolve in the Laves phase, while interface decohesion occurs at the α-Mg/Laves phase interfaces and pores of different shape also form at inclusions. Thus, depending on location and shape of the damage sites, we distinguish three different types in the microstructure (Figure 1): (1) Laves phase cracks are linear shaped groups of black pixels lying on the white Laves phases, (2) interface decohesion at the α-Mg/Laves phase interfaces are black islands lying at the boundary of the white Laves phase and the grey matrix of the Mg, and (3) inclusions are normally randomly shaped and sized with either black, grey, or white pixels lying in the matrix. The pores presumably formed during casting are also placed in the latter category. The Laves phase cracks and interface decohesion sites are of particular interest here as they are deformation induced and not intrinsically dependent on casting or melt conditions.

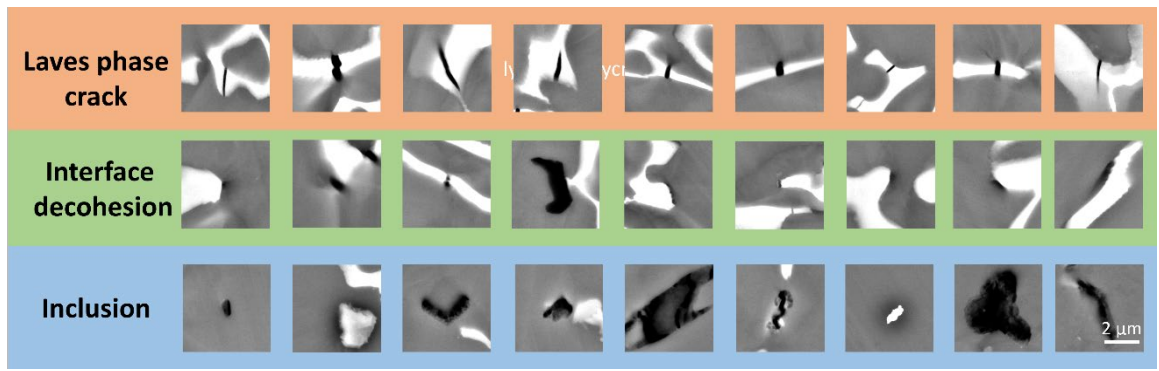

*Figure 1. Illustration of the three types of the damage sites considered in the Mg-Al-Ca microstructure.*

### 2.2.3 Data preparation

As the starting step, a dataset of 452 SEM images collected by both high-resolution scanning electron microscopes with varying sizes of 1024 x 768 and 3072 x 2048 pixels and varying resolutions of 16 and 48 nm/pixel were introduced to the LabelIing software [46]. The three damage site classes, Laves phase cracks, interface decohesion and inclusion, were annotated using the bounding box defining tool by drawing a rectangular box around the damage site and specifying its type. The annotation was performed by a single person after the appearance of each class was defined and agreed between several scientists. As the output, an annotated label file with .xml format is generated and assigned for each SEM image containing the information with respect to coordinates and size of the bounding boxes of the damage sites in the annotated images.

Subsequently, the annotated images together with the respective labels were uploaded to the Roboflow [47] website where the annotation files were converted to YoloV5 format. Thereafter, the data were randomly split into the training and test set with 80-20% proportion respectively (362 training and 90 validation). The size of the bounding box, containing the whole area of the damage site, varied depending on the size of the damage sites in the microstructure. Out of 1082 annotated damage sites in the images of the training dataset, 856, 452 and 143 were annotation as associated with Laves cracks, interface decohesion and inclusions, respectively. The distributions of the different classes depending on the position in the microstructure and more details about the data distribution in the sense of dimensions and coordinates of the bounding boxes are mapped in Figure 2. As can be perceived from the training data distribution maps, there is no specific concentration of the positions of the damage sites within the image, which in turn would introduce no bias to the localization of the damage sites in the model training process and indicates that the selected area is large enough to represent the locally changing microstructure and its variable damage distribution.

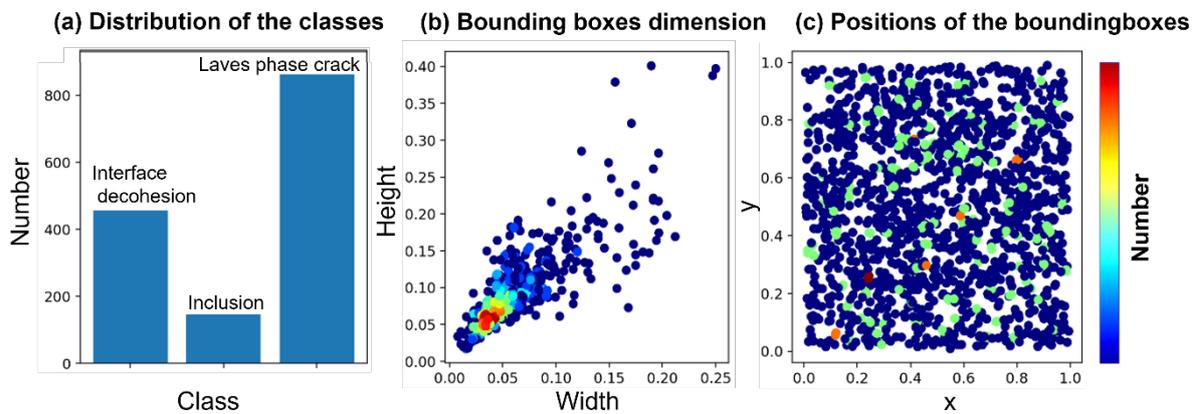

Figure 2. a) The number of the annotation of each class, b) the relative position of the bounding box on an SEM image c) the normalized dimensions of the bounding boxes (the highly occupied coordinates and dimensions are denoted with different colour).

Data augmentation serves as a tool to increase the network robustness against different variations and in turn, enhances the robustness of network to classify the objects (here damage sites) obtained from different environments [38],[48]. The common augmentation method used in the context of YoloV5 is mostly Mosaic data augmentation and colour space augmentation [48]. In Mosaic data augmentation, four training images are mixed. In colour space augmentation, the hue and saturation of the images are varied. In addition, other geometrical augmentation methods like flipping, rotation, translation, scaling which exist in the default settings of the method and do not change the definition of the three classes.

### 2.2.4 Network Training

The training of the network was initiated from scratch, particularly without utilizing pre-existing weights that had been previously trained on an alternate dataset such as COCO [45] (a method known as transfer learning [49]). The rationale behind this decision relates to the higher evaluation metrics, particularly mAPs, which is a comprehensive single indicator summarizing precision, recall and intersection over the union (IoU), when compared to transfer learning methods implemented on the same dataset. This approach has been adopted to ensure a more rigorous and accurate evaluation of the network's performance.

In the YoloV5, the optimal weight is automatically saved based on the best fitness function. We utilised the standard saved weight of the network to perform prediction on our data. By using the best saved weight, we were able to detect and classify damage sites across the previously unseen SEM images, and extract the relevant quantities of each type. In order to assess the performance of the trained network, a number of metrics were employed and are described in detail below.

Precision is calculated as the ratio between the number of *Positive* objects in the image (e.g Laves crack here) correctly classified, to the total number of *Positive* objects (either correctly or incorrectly classified). Equation 1 provides the corresponding formula for calculating the precision [50].

$$Precision = \frac{TP}{TP+FP}$$ *Equation 1*

Recall is calculated as the ratio between the number of *Positive* objects in the image correctly classified, to the number of *Positive* objects correctly classified and *Negative* objects incorrectly classified. It is basically a measure of how well the model finds all the relevant cases in the images (see Equation 2) [51].

$$Recall = \frac{TP}{TP + FN}$$ *Equation 2*

The evaluation of object detection and classification algorithms relies fundamentally on both precision and recall metrics. These metrics are interdependent and exhibit a trade-off relationship. Increasing recall, for example, by generating a greater number of predictions, would result in a corresponding decrease in precision, which refers to the accuracy of prediction. To effectively balance these two metrics, it is necessary to aggregate them into a single metric, which is described in the following.

Average precision (AP) is a parameter calculated by averaging all the precision values across various recall values under different thresholds. The precision is calculated for different classes and, depending on the number of the classes to be detected and classified, several precision-recall relations are built. The threshold utilised is the Intersection Over Union (IoU), which expresses the overlap between the area under the ground truth bounding boxes and the predicted bounding boxes over the whole areas under both bounding boxes (Figure 3) as

$$IoU = \frac{B \cap B^{GT}}{B \cup B^{GT}}$$ *Equation 3*

where $B$ is the area of the predicted bounding box and $B^{GT}$ is the area of ground truth bounding box [52].

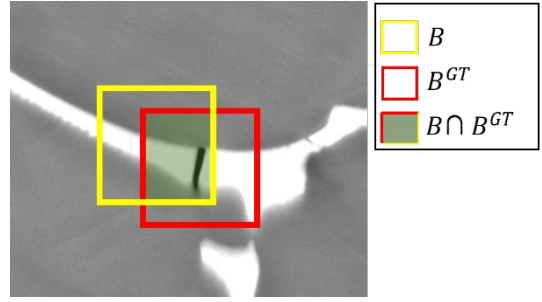

*Figure 3. Illustration of the intersection over union on example case of Laves phase crack.*

Mean average precision (mAP) is the mean value of all AP values across all classes. mAP is the most robust way of evaluating the method. It is mostly used with mAP@0.5 and mAP@0.5:0.95, with the numbers referring to the range of the IoU. mAP is expressed as

$$\text{mAP} = \frac{1}{n}\sum_{k=1}^{k=n} AP_k \qquad \textit{Equation 4}$$

where *n* is the number of the classes, and $AP_k$ is the average precision of each class [53].

### 2.2.5 Evaluation of crack inclination

To study the dependence of the inclination of the formed Laves phase cracks relative to the loading direction, we designed a method based on the application of the Hough transformation. In computer vision, the Hough transform is a widely recognized technique used for the detection of object instances with specific class of shape like lines, circles, or ellipses inside the image [54]. The detection process is conducted through convolutional operations to extract the object boundaries in the image. The collected information can then be used to accumulate Hough votes to finally find the instances of the specific shapes in the image [55].

Our procedure is illustrated in Figure 4. First, the damage sites classified as Laves phase cracks were isolated and cropped to a size of 100 x 100 pixels around the centroid of the object to encompass a single damage site as well as some surrounding area. Sites where more than one crack were contained within the cropped window were omitted. For precise recognition of the crack boundaries, its edges were slightly thickened by means of erosion morphology correction [56] (function parameters: 3*3 kernel size under 1 iteration). Then the edges of the crack were detected by Canny edge detection [57]. Finally, the edge lines of Laves phase cracks were isolated by means of a Hough line transformation. Due to varying brightness and contrast condition of the images, the parameters which influence the voting accumulation, such as intensity threshold values in the canny edge detection as well as ρ-accuracy and vote values of the Hough transformation were dynamically adjusted within the iteration for each image until it captured the lines with the highest number of the votes as the representation of the longest edge line of the crack. In the canny edge detection, the intensity threshold determines the limit to define an edge line as a line with respect to its connectivity, such that if the line portions are connected to the "sure edge" or "non edge" [58] . In the Hough transformation ρ is the distance of each line from the coordination origin and vote is the value of the accumulator [58]. The higher vote here, the higher is the weight which conveys information about the strength of the supporting evidence of the lines [59]. All the packages were implemented using OpenCV library [58]. Finally, the inclination angles of the Laves cracks were calculated in terms of the angle of the isolated parallel lines of the crack edges with respect to the horizontal, along which the tensile stress was applied.

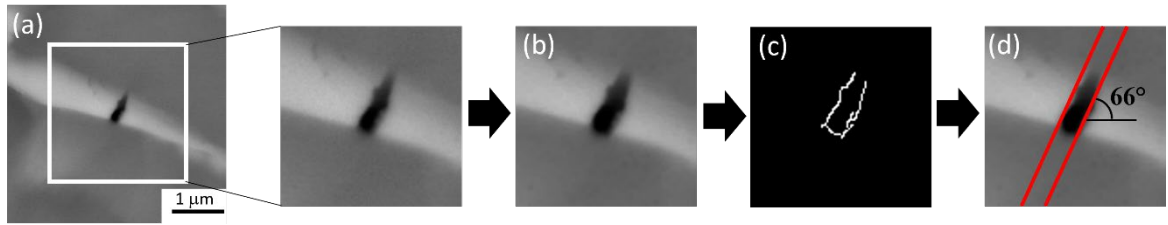

*Figure 4. Illustration of Laves phase crack angle calculation. a) SEM image of the crack cropped from the panorama, b) application of erosion morphology correction, c) Canny edge detection, d) angle calculation.*

## 2.3 Technical Configuration

All computation including training and validation of the network, calculation of numerical results, and detection and classification of damage sites in new images were performed on central high performance computing facilities with a GPU node providing "Nvidia Tesla V100" and 16 GB of memory and 1TB storage. Using this system configuration, training YOLOv5 took around 22 hours. Laves phase crack detection and angle calculation were performed on a workstation with an AMD FX (tm)-8350 eight-core processor and 2 TB of memory.

# 3 Results

## 3.1 Microstructural features of the deformed Mg-4.65Al-2.82Ca alloy

The deformed microstructure of the Mg-465Al-2.82Ca alloy is presented in Figure 5and Figure 6. The darker phase is the α-Mg phase and has the composition 98.8Mg-1.2Al (at. %) as determined by STEM-EDS presented in previous work [39]. The brighter phase is the C36 Ca(Mg,Al)$_2$ Laves phase. A detailed characterisation of this phase and the Mg-C36 interface has already been presented in reference [39]. Deformation in the α-Mg matrix is dominated by basal slip (Figure 5) and extension twinning (Figure 6). Small cracks in the Laves phase tend to nucleate at the points where basal slip lines in α-Mg phase intersect the Laves phase (depicted by red arrows in Figure 5 b). In some instances, slip is also transmitted from α-Mg to the Laves phase (Figure 5 c).

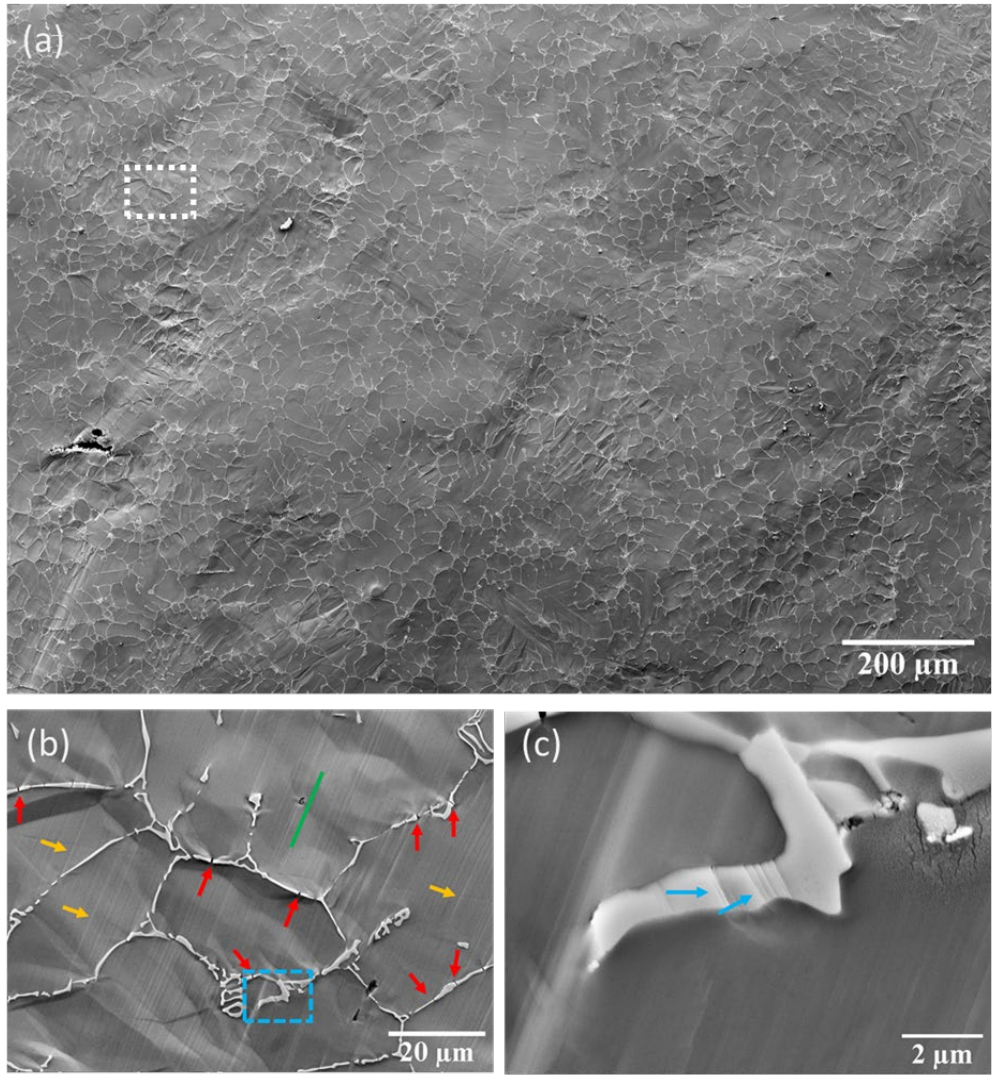

*Figure 5. SE images of 5% deformed Mg-4.65Al-2.82Ca sample, (a) panoramic image of large microstructural area, (b) magnified image of the area enclosed by white rectangle in (a), and (c) magnified portion of the area covered by blue rectangle in (b). Loading direction is horizontal for all images. The basal slip trace is represented by green line in (b).*

Examples of the formation of extension twins in the α-Mg phase are given in Figure 6. The three variants of extension twins within one α-Mg grain are highlighted in Figure 6 a. There is more cracking in Laves phases in α-Mg grains containing significant deformation features, such as slip lines and deformation twins (Figure 6 a-b). On the other hand, grains which do not show these deformation features appear to contain fewer cracks in the Laves phase. (Figure 6 c). The cracks highlighted by red arrows (Figure 6 d and e) have the same orientation as those of the slip lines highlighted by blue arrows in the C36 Laves phase. This indicates that the cracks may have appeared in the Laves phase after plastic deformation through dislocation slip. Moreover, as the strain rate during deformation decreases, other deformation features, such as α-Mg/Laves phase interface decohesion, begin to appear more frequently (Figure 6 f). This shows that the type of damage initiated in the Mg-4.65Al-2.82Ca alloy is rate dependent.

By this selective and manual analysis, we therefore find that the type of damage and inclination of Laves phase cracks depends on the local microstructural and deformation conditions. To investigate this further, a statistical analysis is needed, which is provided in the following using a combination of artificial intelligence and classical image analysis.

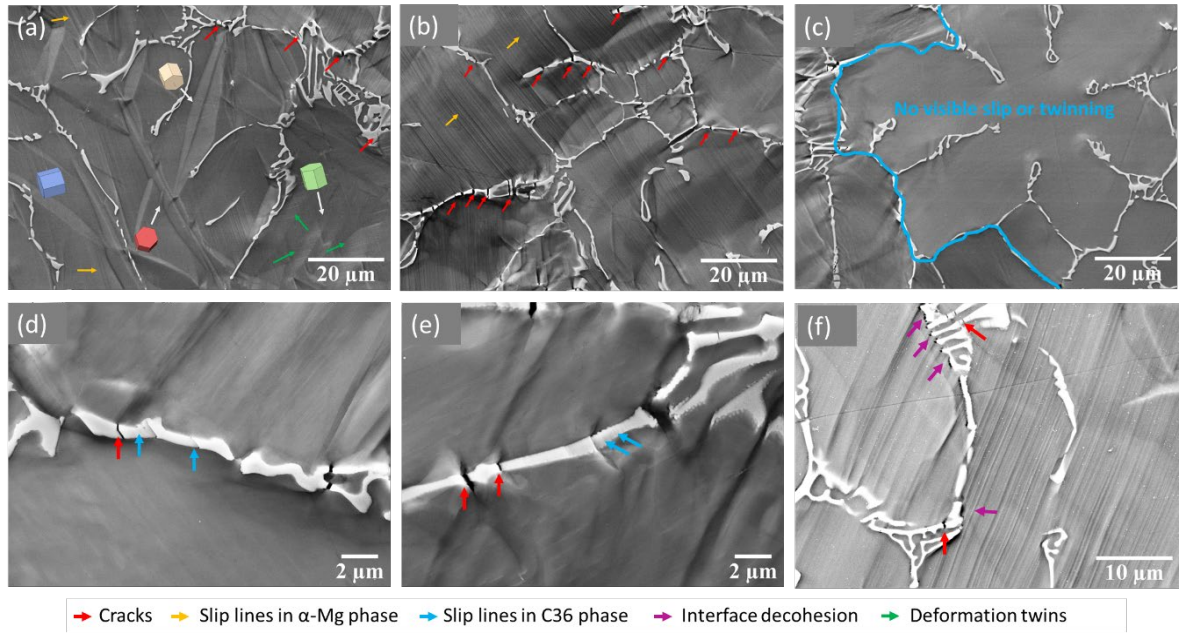

*Figure 6: SE images of 5% deformed Mg-4.65Al-2.82Ca sample at a strain rate of $5·10^{-4}$/s, (a) three variants of extension twins visible in same α-Mg grain, (b) slip lines in α-Mg phase, c) α-Mg grain bounded by blue lines shows no evidence of slip or twinning, (d-e) cracks in the Laves phase as depicted by red arrows and slip lines by blue arrows. The orientation of cracks in C36 phase is same as that of slip traces in C36 phase. (f) microstructure of sample deformed to 5% strain at a strain rate of $5·10^{-6}$/s*

## 3.2 Damage analysis using deep learning

### 3.2.1 Network training

The history of network training in terms of precision, recall and mean average precision at two IoU thresholds of 0.5 and 0.5 to 0.95 are depicted in the graphs in Figure 7. The highest mean average precision (mAP) at the intersection over union in the range of 0:5 to 0.95 reaches 70%.

The batch size and number of epochs were 4 and 2500 respectively. Batch sizes of 2 and 8 were also explored, however, the smaller batch size slowed down the training process and the higher one resulted in a reduction of the generalizability of the model and also increased the memory usage. The number of the epochs was fixed to 2500 as a compromise between increasing trend of network learning potential and overfitting.

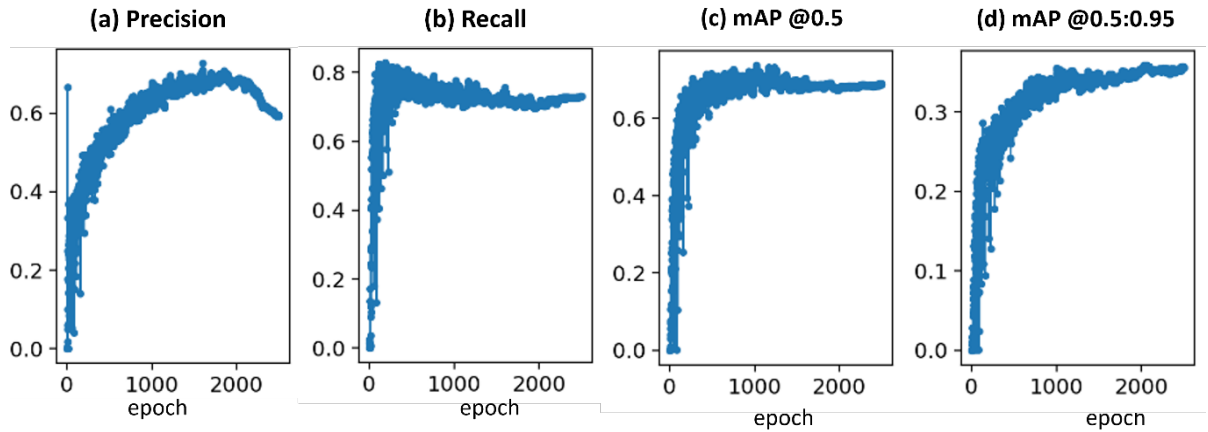

*Figure 7. Evaluation metrics during training history of the network*

### 3.2.2 Rate dependence of damage formation

With decreasing strain rates from $5\cdot10^{-4}$/s to $5\cdot10^{-6}$/s, the relative fraction of the observed damage mechanisms was found to change significantly. Both samples presented in Figure 8 and Figure 9 were deformed at 170 °C to 5% global strain. The sample deformed at a higher strain rate ($5\cdot10^{-4}$/s) exhibits predominantly cracks running through the Laves phase struts (see Figure 8). The frequency of red squares and arrows (depicting cracking in Laves phase) is much higher than the green boxes and arrows (enclosing cracks at the α-Mg/Laves phase interfaces).

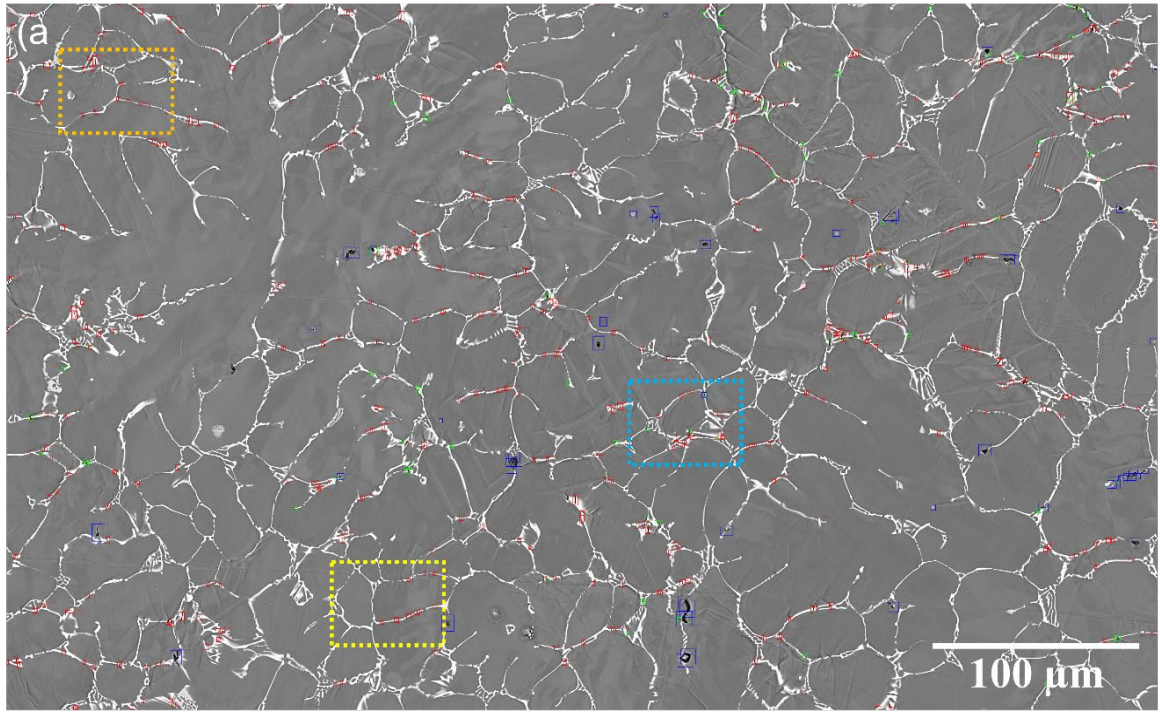

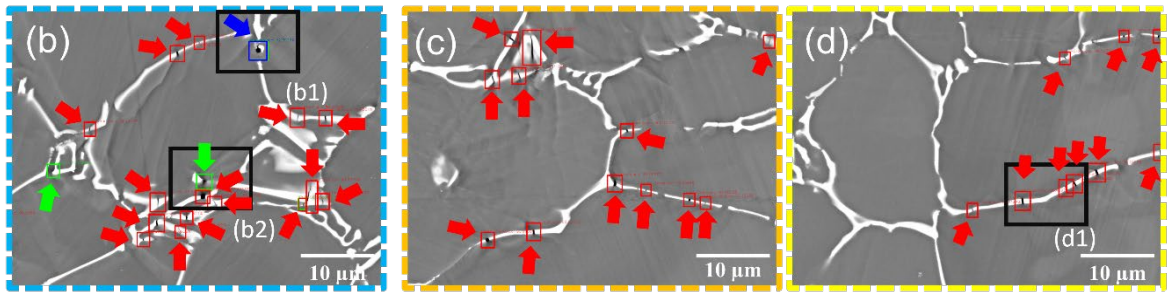

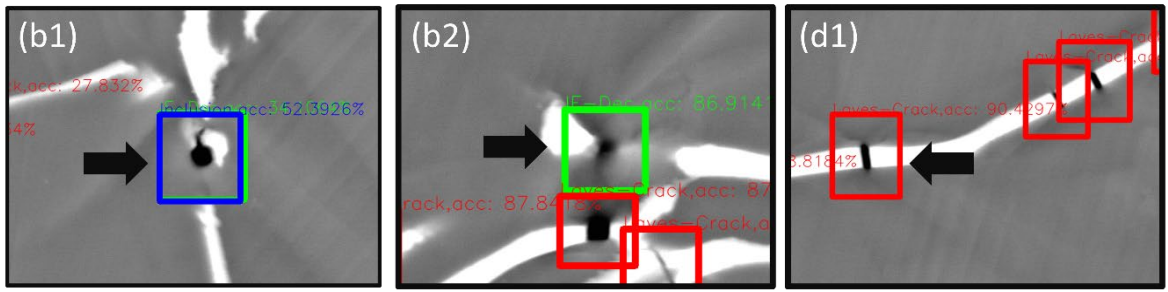

*Figure 8. (a) and (b): BSE panorama from a sample surface deformed upto 5 % global strain at a strain rate of $5·10^{-4}$/s and temperature of 170 °C. Magnified images of the microstructural regions enclosed by the blue, orange and yellow rectangles are presented in (b), (c), and (d), respectively. Green boxes highlight interface decohesion site at α-Mg/Laves phase interfaces, red corresponds to cracks in the Laves phase, while blue boxes represent the inherent defects in the material visible on sample surface, such as inclusions and other pores.*

Much more interface decohesion, as highlighted by green squares and arrows, was observed in samples deformed at a strain rate of $5·10^{-6}$/s (compare Figure 8 and Figure 9). Although Laves phase cracking is also visible in the sample deformed at the lower strain rate of 5·10⁻⁶/s, the ratio between Laves phase cracking and interface decohesion is lower compared to the sample deformed at higher strain rate.

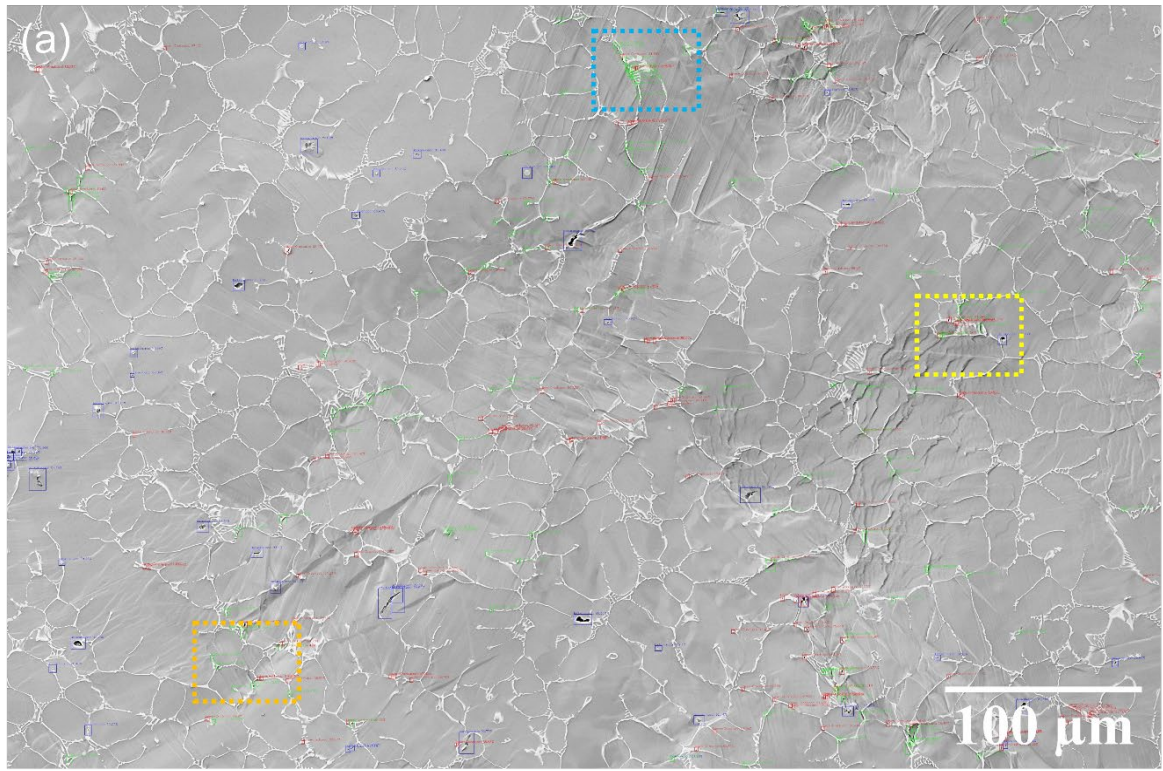

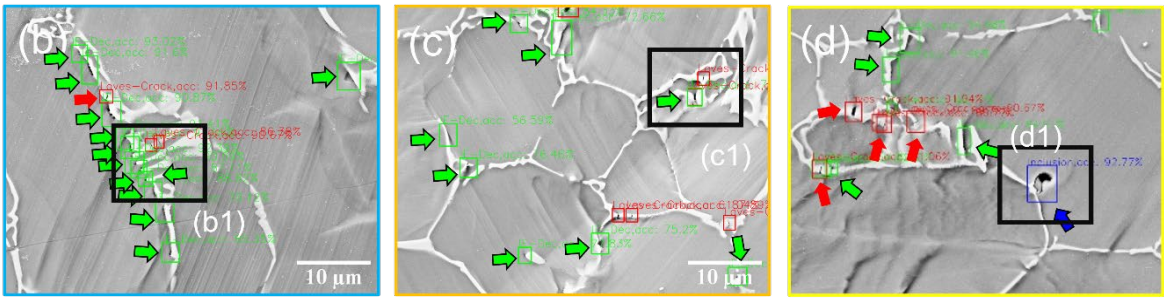

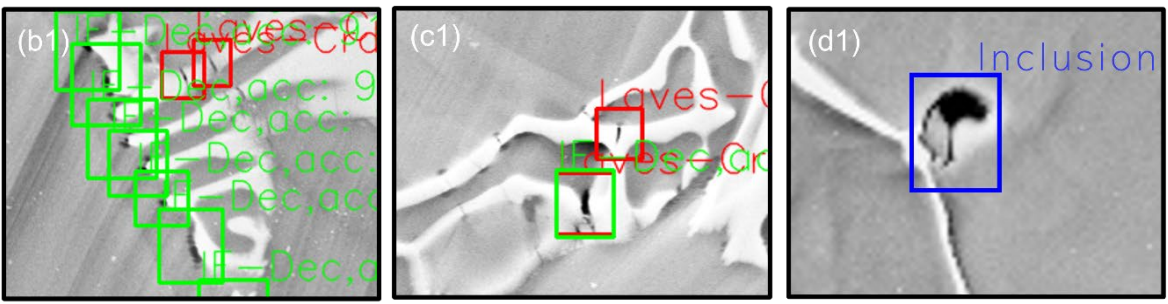

*Figure 9.(a): SE panorama from a sample surface deformed upto 5 % global strain at a strain rate of 5·10⁻⁶/s and temperature of 170 °C. Magnified images of the microstructural regions enclosed by the blue, red and yellow rectangles are presented in (b), (c), and (d). Green boxes highlight interface decohesion sites at α-Mg/Laves phase interfaces, red indicates the cracks in the Laves phase, while blue boxes represent inclusions and other pores.*

The quantitative data for all experiments is presented in Figure 10.

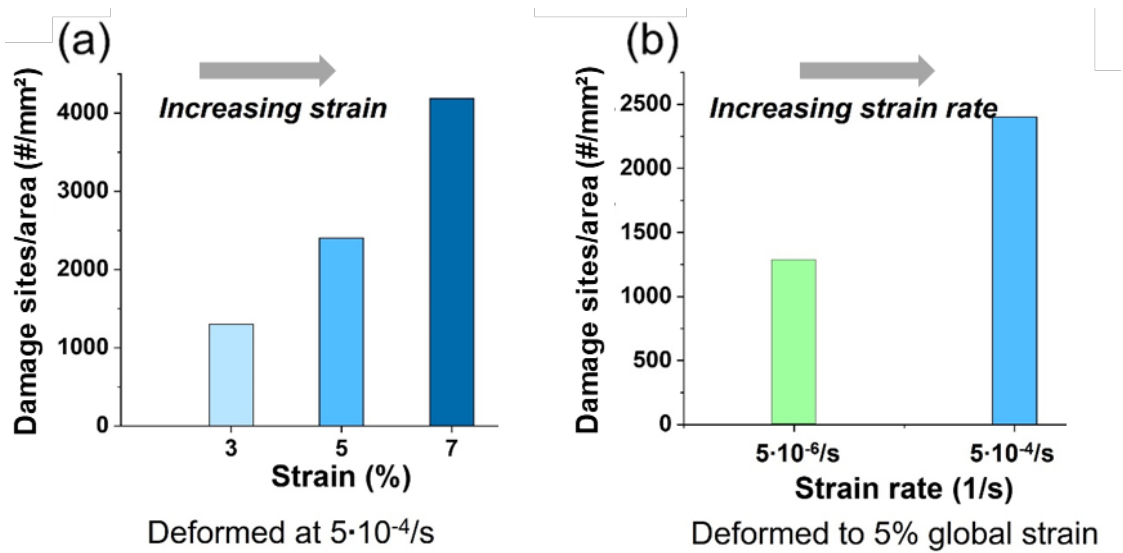

*Figure 10. Data calculated using AI from the Mg-4.65Al-2.82Ca deformed at a temperature of 170 °C. (a) Total number of damage sites per area in the alloy when subjected to different strain levels at the same rate, (b) change in total number of damage sites per area for the two different strain rates.*

As expected, the total number of damage sites per area for a given strain rate varied directly with strain Figure 10 a. Higher strain resulted in an increased density of damage sites. At all three strain levels and the higher strain rate of $5·10^{-4}$/s, the fraction of Laves cracks exceeded that of interface decohesion sites by a factor of two or more. However, the nucleation of damage sites reduced significantly as the strain rate was reduced from $5·10^{-4}$/s to $5·10^{-6}$/s for the same global strain of 5%. Further, the relative fraction of the two deformation induced damage types changed with strain rate ( Figure 11). Interface decohesion is a significant mode of damage at a lower strain rate ($5·10^{-6}$/s), accounting for approximately half the total number of damage sites, while at high strain rate ($5·10^{-4}$/s), cracks in the Laves phase dominate.

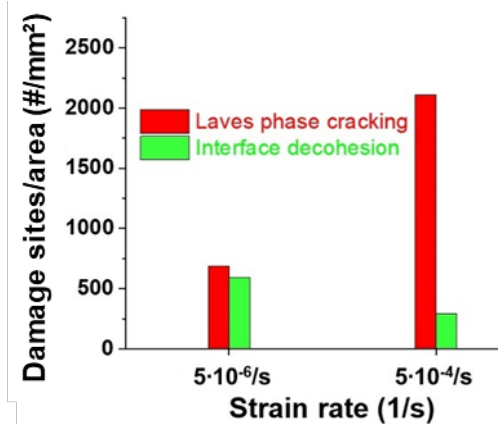

*Figure 11. Change in Laves phase crack and interface decohesion fraction as a function of strain rate in the Mg-4.65Al-2.82Ca alloy deformed at 170 °C to 5% global strain.*

# 4 Discussion

## 4.1 4.1 Microstructural and mechanical heterogeneity in the Mg-Al-Ca alloy

During straining, deformation therefore usually initiates in the α-Mg phase [8, 28] and then it either extends into the Laves phase (Figure 5c and Figure 6d,e) or creates enough local stress to initiate cracks in the Laves phase (Figure 5b and Figure 6). This observation is consistent with previous experimental and computational work [8, 28] and result from the differences in mechanical properties and crystal structure between α-Mg and Laves phases [27]. In its as-cast form, the Mg-4.65Al-2.82Ca alloy has a microstructure consisting of α-Mg phase and C36 Laves phase. The α-Mg phase in Mg-Al-Ca alloys is a solid solution primarily of Al in Mg [3, 8, 12, 27, 28, 60]. This phase predominantly deforms via basal slip and deformation twinning [8, 12, 27, 28]. In pure Mg, these two deformation mechanisms have the lowest critical resolved shear stresses (CRSS) of < 1 MPa and < 10 MPa respectively at the macroscopic scale [61-64]. In contrast, the CRSS for prismatic and pyramidal slip are of the order of 40 MPa [65-67]. Laves phases on the other hand demonstrate much higher hardness and strength but extreme brittleness at low temperatures [13-15]. However, these phases show plasticity at small scales even at room temperature, for example, in micropillar compression [6, 16, 68, 69]. The CRSS values for basal, prismatic, and pyramidal slip values in the C14 ($CaMg_2$, structurally quite close to the C36 Laves phase) phase were found to be of the order of 0.5 GPa [6]. These values decrease with temperature but they remain well above the CRSS of Mg even at high temperature of the order of 250 °C [5]. Similarly, the CRSS values observed for $\{111\}(1\bar{1}0)$ of C15 ($CaAl_2$) Laves phase was reported to be nearly ten times to that observed in α-Mg phase [16].

In addition to this generally heterogeneous deformation resulting from the mechanical contrast, we also observed that cracking in the Laves phase is more concentrated in those α-Mg grains that exhibited a higher amount of basal slip and extension twinning (see panoramic images in Figure 1 and Figure 2).

This is again in agreement with earlier work on a similar alloy, where it was shown that cracks initiate in the Laves phase either at places where slip lines or twins in the α-Mg phase intersect with the Laves phase [27]. The behaviour of the alloy investigated here is therefore in good agreement with previous reports in the literature. However, the effect of strain and in particular rate has not been investigated explicitly to our knowledge, although a transition from cracking to plastic co-deformation has been predicted by atomistic simulations for an α-Mg-CaMg$_2$ C14 Laves interface [28].

## 4.2 Quantitative damage analysis by deep learning

We used Yolo5s as a single-stage target recognition algorithm. This method proved simple and more effective compared to other deep learning object detection and classification methods like R-CNN, which is more complex and computationally intensive. Owing to its single-shot approach, it is well suited for the detection of small objects, like fine damage sites in our case [44].

The classification performance of the trained network is visualised in Figure 12 for two randomly selected unseen sets of image data using confusion matrices in which actual damage type (from manually labelled ground truth data) and predicted damage type (by the network) are compared. The relatively high scores show that the classification is quite reliable, in particular for the classification of Laves cracking, while the classification of interface decohesion and inclusion sites is more variable. This may be improved by providing more labelled training data, further image augmentation or by including image artefacts or further damage types as additional classification options [38].

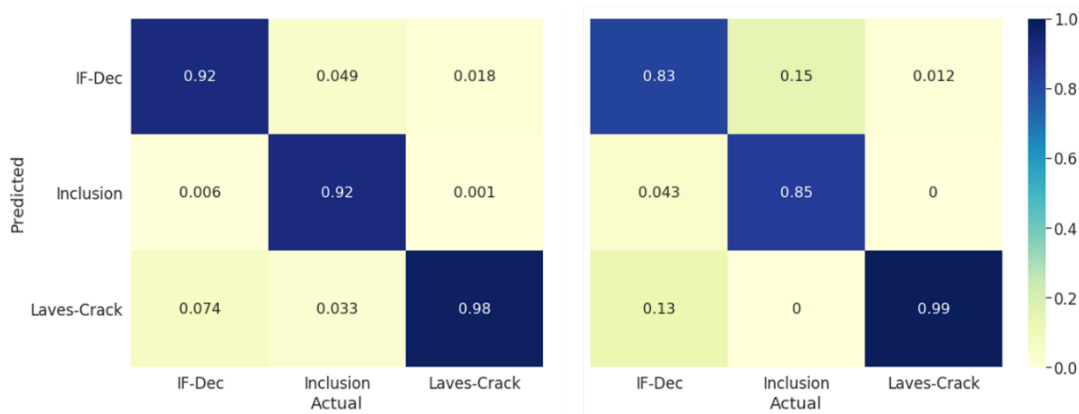

*Figure 12. Confusion matrices of classification associated with 2 randomly selected datasets.*

## 4.3 Strain dependence of damage

The number of deformation induced damage sites formed in the Mg-4.65Al-2.82 Ca alloy increased with strain (Figure 10 a). This is as expected as higher strains will result in more dislocation pileups at α-Mg/Laves phase interfaces and thus more interface decohesion and cracking at and in the intermetallic phase [8]. In polycrystalline Ti-6Al with low interstitial content, Huang et al. [70] have shown that the geometric incompatibility and lack of slip transfer across grain boundaries are mainly responsible for nucleation of microcracks at these boundaries. Significant stress concentrations at grain boundaries can alternatively be released by slip transfer across boundaries, resulting in more homogenous plasticity in the sample [70, 71]. However, in the material investigated here, the dominant interface is a phase boundary rather than grain boundary and the crystal structure of the C36 Laves phase [72] is very different from that of α-Mg phase in spite of both possessing a hexagonal unit cell. Moreover, a dominant orientation relationship found for such alloys places the basal plane of the C36 Laves phase at an approximately perpendicular angle to the basal plane of the α-Mg phase [8, 28]. The different crystal structure and lattice parameters [72] of C36 Laves phase, together with a much higher CRSS [5, 6] for basal slip, as compared to the Mg phase, and unfavourable orientation relationship [8, 28] with Mg therefore severely restricts basal to basal slip transfer into the C36 phase. Slip could possibly transfer when non basal slip is initiated in regions close to α-Mg/C36 Laves phase interfaces as a result of the high and multiaxial stresses induced by a basal dislocation pileup [8].

Consistent alignment of slip traces indicating slip transfer at high angles between the slip planes has indeed been found here (blue arrows in Figure 5 and Figure 6). However, cracks are observed much more commonly, and we therefore could not yet investigate the conditions for slip transfer statistically using the employed neural networks. There may also be a transition from slip in the C36 phase to crack opening along the slip plane, as has been observed in micropillars of similarly complex intermetallic phases [73]. We consider this in more detail below as part of the rate dependence of damage formation, as decohesion of slip planes within the Laves phase would be expected to depend on the extent of slip transmission at the interface and dislocation motion in the intermetallic, both of which are expected to be thermally activated and therefore rate dependent.

### 4.4 Rate dependence of damage

The variation of strain rate resulted in a more varied effect on the formation of damage sites. The total number of sites per area for the same strain of 5% reduced to nearly half as the strain rate was decreased by two orders of magnitude, from $5 \cdot 10^{-4}$/s to $5 \cdot 10^{-6}$/s (Figure 10 b). Moreover, the type of damage changed from predominantly Laves phase cracking to a combination of interface decohesion and Laves phase cracking Figure 11. This change in dominant damage mechanism is likely due to thermally activated phenomena at the interfaces.

In Figure 13, the angle of the Laves phase cracks relative to the loading direction is shown for both strain rates. The angles were calculated for all sites detected and classified by the neural network with sufficient crack spacing to apply the Hogh transformation as described above and benefits from the reliable classification of the Laves phase cracks(Figure 12). The results indicate that there are in fact likely three aspects to consider: in addition to (1) brittle cracking of the Laves phase driven by normal stresses and (2) interface decohesion, (3) Laves phase fracture following slip in the Laves phase or at least introduction of a critical dislocation density near the interface in the Laves phase may occur. Only the latter may be expected to lead to the change in dominant angle away from the perpendicular orientation at 90°, as found for the higher strain rate (Figure 13). In contrast, at the lower rate, the distribution does not show a clear trend and if the data are interpreted to contain a maximum, then it is at a much lower angle. Additional studies on the errors associated with angle distribution measurements relating to, for example, orientation relationships and texture, are needed to investigate whether a preferential angle does exist and , if so, how it is determined.

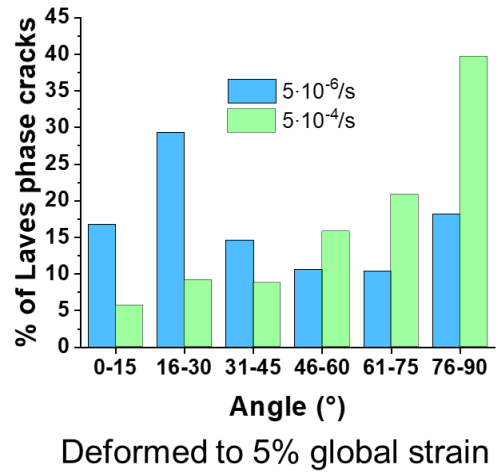

*Figure 13. Angle of damage sites nucleated in the Laves phase relative to the loading direction at the two different strain rates. Sites perpendicular to the loading direction are represented by an angle of 90°, while those parallel to the loading direction are represented by an angle of 0°.*

Purely brittle failure should be independent of thermal activation in the absence of thermally activated mechanisms promoting critical defect formation or pronounced changes in the crystal's stiffness and decohesion energy. We therefore consider here the formation of perpendicular cracks as not thermally activated as a first estimate and discuss in the following how thermal activation may affect the formation of both interface decohesion and non-perpendicular cracks.

Cracks at angles significantly away from 90° to the loading axis are likely to result in large number not simply from local variations of the macroscopic stress field due to the skeleton, but mainly from the formation of stress concentrations, crack nuclei and crystal planes with low decohesion energy in the Laves phase. The first will result directly from the intersection of slip bands and twins in the metallic matric with the intermetallic, whereas the deformation induced initiation of crack nuclei and dislocations in the Laves phase that may lower decohesion energy, will form subsequently at stress concentrations. In combined experimental and computational work on a comparable system, using a similar alloy in micromechanical testing and TEM experiments and a Mg/CaMg$_2$ composite in correlated atomistic simulations, evidence of plastic co-deformation between Mg and Laves phase, interface sliding and fracture has also been observed, consistent with our experimental findings, as shown for example in Figure 5. In case of slip transfer for plastic co-deformation, the active planes are very limited in both phases with <a> dislocations moving on the basal plane of Mg and the basal and prismatic planes of the Laves phase. The dominant orientation relationships identified for other Mg-Al-Ca alloys are based on parallel [22] and near perpendicular basal plane orientations [28, 74], with the latter observed in alloys with very similar processing and microstructure to those investigated here. This implies a strong change in slip plane orientation where co-deformation takes place, which is thought to be enabled by the formation of non-basal dislocations in the Mg matrix as a result of the stress field of the pile-up of basal dislocations at the interface [8]. A likely case of this kind of plastic co-deformation is pictured in Figure 5c. Plastic deformation in the Laves phase is strongly localised with large strain accommodated on individual parallel slip planes. This is typical and in the structurally related μ-phase has been shown to coincide with decohesion of planes on which plasticity led to a high dislocation density [5, 6, 68, 73, 75]. The latter gives a route from plastic co-deformation to the observation of cracks along slip planes in the Laves phase. Atomistic simulations considering Mg/Laves interfaces suggest that the slip transfer may be thermally activated [8, 28] and therefore expected to become more frequent as the temperature is increased or the rate is lowered. Similarly, the motion of dislocations in the Laves phase after slip transfer or nucleation from a stress concentration may be thermally activated, although the extent of this thermal activation at up to 170 °C may be very limited, as the hardness of the related $CaMg_2$ C14 Laves phase has been found to be constant between room temperature and 250 °C [5] or drop only a little up to 60% of its melting temperature (at 320 °C) [9]. Only limited data exist on individual slip systems from microcompression at elevated temperature [5].

Possible mechanisms that give rise to damage classified here as interface decohesion have also been considered previously in experiments and computational studies [8, 28]. In the case of interfacial sliding, it was shown in atomistic simulations that dislocations that pile up against the intermetallic are absorbed into the interface [28]. In nanomechanical testing, a clear signature of thermally activated deformation was observed, however, as both sliding and slip transfer were found to occur undeath indentations [8], a direct correlation with either mechanism was not possible by indentation.

The approach presented here now adds valuable additional insights. It allows us to study the prevalence of the individual mechanisms and how their competition is affected by applied strain and strain rate. In particular, it has revealed a transition in active mechanism and, within the observed class of cracks in the Laves phases, important clues also towards changing mechanisms of plastic co-deformation and their impact on crack formation. These insights were found to be consistent with previous experimental and computational studies on very similar alloys and related strengthening Laves phases or idealised by directly relevant phase boundaries. In future experiments, additional parameters may now be considered, especially deformation temperature, the morphology of the intermetallic skeleton and the local orientation relationships. The intermetallic strut size, volume fraction and present Laves phase can be controlled by alloy composition, cooling rate after casting and subsequent annealing [74]. Concurrent orientation imaging by EBSD across large areas at sufficiently low kV to successfully index the small Laves phase volumes, will rely heavily on appropriate metallographic preparation and indexation method [12, 76]. Digital image correlation and surface topography analysis for (quasi) in-situ experiments may additionally help to reveal the role of interfacial sliding both in and normal to the surface plane and it may be possible to integrate this data into a workflow with automated damage type analysis.

# 5 Conclusions

We investigated the changes in deformation induced damage density and type with strain and strain rate at 170 °C in a Mg-4.65Al-2.82Ca alloy consisting of an interconnected C36 Laves phase skeleton embedded in an α-Mg matrix. The main conclusions from this work are

- at higher rate ($5 \cdot 10^{-6}$/s), damage formation is dominated by cracking in the Laves phase (≈88% of all deformation induced damage sites),
- at lower rate ($5 \cdot 10^{-4}$/s), a transition towards interface decohesion (≈46% vs 54% Laves phase cracking) is observed
- the orientation of the formed cracks in the Laves phase changes from predominant cracking perpendicular to the loading axis at the higher rate to much more random crack inclination, which we associate with beginning plastic co-deformation across the interfaces.
- In the two phase microstructure spanning microstructural length scales from single to hundreds of μm, these insights were made possible based on high resolution data from an area greater than 4mm$^2$. The use of deep learning for damage detection and classification as well as subsequent image analysis proved successful as well as essential in analysing this large image dataset.

Future work may build on these insights and methods to further unravel the role of thermal activation at interface boundaries and the role of plasticity in the reinforcing Laves phase, which may facilitate greater ductility of the alloy by plastic co-deformation on the one hand or lead to increased microscopic damage formation on the other.

# 6 Acknowledgement

The authors are grateful for the financial support received from the Deutsche Forschungsgemeinschaft (DFG) as part of Collaborative Research Center CRC 1394 - Structural and Chemical Atomic Complexity: From Defect Phase Diagrams to Material Properties (project ID 409476157) and Collaborative Research Center TRR 188 – Damage controlled forming (project ID 278868966). This work is also funded by the state of North Rhine-Westphalia as part of the NHR Program. Calculations

# 7 References

[1] D. Amberger, P. Eisenlohr, M. Goken, On the importance of a connected hard-phase skeleton for the creep resistance of Mg alloys, Acta Materialia 60(5) (2012) 2277-2289.
[2] D. Amberger, P. Eisenlohr, M. Göken, Microstructural evolution during creep of Ca-containing AZ91, Materials Science and Engineering: A 510-511 (2009) 398-402.
[3] M. Zubair, S. Sandlöbes, M.A. Wollenweber, C.F. Kusche, W. Hildebrandt, C. Broeckmann, S. Korte-Kerzel, On the role of Laves phases on the mechanical properties of Mg-Al-Ca alloys, Materials Science and Engineering: A 756 (2019) 272-283.
[4] H.N. Mathur, V. Maier-Kiener, S. Korte-Kerzel, Deformation in the γ-$Mg_{17}Al_{12}$ phase at 25–278 °C, Acta Materialia 113 (2016) 221-229.
[5] M. Freund, D. Andre, C. Zehnder, H. Rempel, D. Gerber, M. Zubair, S. Sandlöbes-Haut, J.S.K.L. Gibson, S. Korte-Kerzel, Plastic deformation of the $CaMg_2$ C14-Laves phase from 50 - 250°C, Materialia 20 (2021) 101237.
[6] C. Zehnder, K. Czerwinski, K.D. Molodov, S. Sandlöbes-Haut, J.S.K.L. Gibson, S. Korte-Kerzel, Plastic deformation of single crystalline C14 $Mg_2Ca$ Laves phase at room temperature, Materials Science and Engineering: A 759 (2019) 754-761.
[7] M. Freund, D. Andre, P.L. Sun, C.F. Kusche, S. Sandlöbes-Haut, H. Springer, S. Korte-Kerzel, Plasticity of the C15-$CaAl_2$ Laves phase at room temperature, Materials & Design 225 (2023) 111504.
[8] M. Zubair, S. Sandlöbes-Haut, M. Lipińska-Chwałek, M.A. Wollenweber, C. Zehnder, J. Mayer, J.S.K.L. Gibson, S. Korte-Kerzel, Co-deformation between the metallic matrix and intermetallic phases in a creep-resistant Mg-3.68Al-3.8Ca alloy, Materials & Design 210 (2021) 110113.
[9] C. Kirsten, P. Paufler, G.E.R. Schulze, Zur plastischen Verformung intermetallischer Verbindungen, Monatsberichte der Deutschen Akademie der Wissenschaften 6 (2) (1964) 140-147.
[10] T.H. Müllerr, P. Paufler, Yield strength of the monocrystalline intermetallic compound $MgZn_2$, 40(2) (1977) 471-477.
[11] P. Paufler, Early work on Laves phases in East Germany, Intermetallics 19(4) (2011) 599-612.
[12] D. Andre, M. Freund, U. Rehman, W. Delis, M. Felten, J. Nowak, C. Tian, M. Zubair, L. Tanure, L. Abdellaoui, H. Springer, J.P. Best, D. Zander, G. Dehm, S. Sandlöbes-Haut, S. Korte-Kerzel, Metallographic preparation methods for the Mg based system Mg-Al-Ca and its Laves phases, Materials Characterization 192 (2022) 112187.
[13] F. Stein, A. Leineweber, Laves phases: a review of their functional and structural applications and an improved fundamental understanding of stability and properties, Journal of Materials Science 56(9) (2020) 5321-5427.
[14] C.T. Liu, J.H. Zhu, M.P. Brady, C.G. McKamey, L.M. Pike, Physical metallurgy and mechanical properties of transition-metal Laves phase alloys, Intermetallics 8(9) (2000) 1119-1129.
[15] A. Von Keitz, G. Sauthoff, Laves phases for high temperatures—Part II: Stability and mechanical properties, Intermetallics 10(5) (2002) 497-510.
[16] S. Luo, L. Wang, J. Wang, G. Zhu, X. Zeng, Micro-compression of $Al_2Ca$ particles in a Mg-Al-Ca alloy, Materialia (2021) 101300.
[17] N. Takata, H. Ghassemi Armaki, Y. Terada, M. Takeyama, K.S. Kumar, Plastic deformation of the C14 Laves phase $(Fe,Ni)_2Nb$, Scripta Materialia 68(8) (2013) 615-618.
[18] A.V. Kazantzis, M. Aindow, I.P. Jones, G.K. Triantafyllidis, J.T.M. De Hosson, The mechanical properties and the deformation microstructures of the C15 Laves phase Cr2Nb at high temperatures, Acta Materialia 55(6) (2007) 1873-1884.
[19] H.A. Elamami, A. Incesu, K. Korgiopoulos, M. Pekguleryuz, A. Gungor, Phase selection and mechanical properties of permanent-mold cast Mg-Al-Ca-Mn alloys and the role of Ca/Al ratio, J Alloy Compd 764 (2018) 216-225.

[20] S. Sanyal, M. Paliwal, T.K. Bandyopadhyay, S. Mandal, Evolution of microstructure, phases and mechanical properties in lean as-cast Mg–Al–Ca–Mn alloys under the influence of a wide range of Ca/Al ratio, Materials Science and Engineering: A 800 (2021) 140322.
[21] C. Ma, W. Yu, X. Pi, A. Guitton, Study of Mg–Al–Ca magnesium alloy ameliorated with designed Al8Mn4Gd phase, Journal of Magnesium and Alloys 8(4) (2020) 1084-1089.
[22] A.A. Luo, M.P. Balogh, B.R. Powell, Creep and microstructure of magnesium-aluminum-calcium based alloys, Metall Mater Trans A 33(3) (2002) 567-574.
[23] S.M. Zhu, B.L. Mordike, J.F. Nie, Creep properties of a Mg–Al–Ca alloy produced by different casting technologies, Materials Science and Engineering: A 483-484 (2008) 583-586.
[24] B. Kondori, R. Mahmudi, Effect of Ca additions on the microstructure, thermal stability and mechanical properties of a cast AM60 magnesium alloy, Materials Science and Engineering: A 527(7) (2010) 2014-2021.
[25] Y. Nakaura, A. Watanabe, K. Ohori, Effects of Ca,Sr Additions on Properties of Mg-Al Based Alloys, MATERIALS TRANSACTIONS 47(4) (2006) 1031-1039.
[26] B.R. Powell, P.E. Krajewski, A.A. Luo, Magnesium alloys for lightweight powertrains and automotive structures, Materials, Design and Manufacturing for Lightweight Vehicles2021, pp. 125-186.
[27] M. Zubair, S. Sandlöbes-Haut, M.A. Wollenweber, K. Bugelnig, C.F. Kusche, G. Requena, S. Korte-Kerzel, Strain heterogeneity and micro-damage nucleation under tensile stresses in an Mg–5Al–3Ca alloy with an intermetallic skeleton, Materials Science and Engineering: A 767 (2019).
[28] J. Guénolé, M. Zubair, S. Roy, Z. Xie, M. Lipińska-Chwałek, S. Sandlöbes-Haut, S. Korte-Kerzel, Exploring the transfer of plasticity across Laves phase interfaces in a dual phase magnesium alloy, Materials & Design (2021) 109572.
[29] S.W. Xu, N. Matsumoto, K. Yamamoto, S. Kamado, T. Honma, Y. Kojima, High temperature tensile properties of as-cast Mg–Al–Ca alloys, Materials Science and Engineering: A 509(1) (2009) 105-110.
[30] G. Zhang, K.-q. Qiu, Q.-c. Xiang, Y.-l. Ren, Creep resistance of as-cast Mg-5Al-5Ca-2Sn alloy, China Foundry 14(4) (2017) 265-271.
[31] G. Zhu, L. Wang, J. Wang, J. Wang, J.-S. Park, X. Zeng, Highly deformable Mg–Al–Ca alloy with $Al_2Ca$ precipitates, Acta Materialia 200 (2020) 236-245.
[32] H.J. Frost, M.F. Ashby, Deformation-Mechanism Maps, The Plasticity and Creep of Metals and Ceramics, Franklin Book Company Inc.1982.
[33] R. Yamashita, M. Nishio, R.K.G. Do, K. Togashi, Convolutional neural networks: an overview and application in radiology, Insights into imaging 9 (2018) 611-629.
[34] R. Girshick, J. Donahue, T. Darrell, J. Malik, Rich feature hierarchies for accurate object detection and semantic segmentation, Proceedings of the IEEE conference on computer vision and pattern recognition, 2014, pp. 580-587.
[35] Y. Li, X. Feng, Y. Liu, X. Han, Apple quality identification and classification by image processing based on convolutional neural networks, Scientific Reports 11(1) (2021) 16618.
[36] Q. Li, W. Cai, X. Wang, Y. Zhou, D.D. Feng, M. Chen, Medical image classification with convolutional neural network, 2014 13th international conference on control automation robotics & vision (ICARCV), IEEE, 2014, pp. 844-848.
[37] C. Kusche, T. Reclik, M. Freund, T. Al-Samman, U. Kerzel, S. Korte-Kerzel, Large-area, high-resolution characterisation and classification of damage mechanisms in dual-phase steel using deep learning, PloS one 14(5) (2019) e0216493.
[38] S. Medghalchi, C.F. Kusche, E. Karimi, U. Kerzel, S. Korte-Kerzel, Damage Analysis in Dual-Phase Steel Using Deep Learning: Transfer from Uniaxial to Biaxial Straining Conditions by Image Data Augmentation, JOM 72(12) (2020) 4420-4430.
[39] M. Zubair, M. Felten, B. Hallstedt, M. Vega Paredes, L. Abdellaoui, R. Bueno Villoro, M. Lipinska-Chwalek, N. Ayeb, H. Springer, J. Mayer, B. Berkels, D. Zander, S. Korte-Kerzel, C. Scheu, S. Zhang, Laves phases in Mg-Al-Ca alloys and their effect on mechanical properties, Materials & Design 225 (2023) 111470.
[40] Image Composite Editor. https://www.microsoft.com/en-us/research/product/computational-photography-applications/image-composite-editor/.

[41] Glenn Jocher, Ayush Chaurasia, Alex Stoken, Jirka Borovec, NanoCode012, Yonghye Kwon, Kalen Michael, TaoXie, J. Fabg, ultralytics/yolov5: v7.0 - YOLOv5 SOTA Realtime Instance Segmentation, 2022. https://doi.org/10.5281/zenodo.7347926. 2022).
[42] J. Redmon, S. Divvala, R. Girshick, A. Farhadi, You only look once: Unified, real-time object detection, Proceedings of the IEEE conference on computer vision and pattern recognition, 2016, pp. 779-788.
[43] M. Wang, B. Yang, X. Wang, C. Yang, J. Xu, B. Mu, K. Xiong, Y. Li, YOLO-T: Multitarget Intelligent Recognition Method for X-ray Images Based on the YOLO and Transformer Models, Applied Sciences 12(22) (2022) 11848.
[44] M. Horvat, G. Gledec, A comparative study of YOLOv5 models performance for image localization and classification, Central European Conference on Information and Intelligent Systems, Faculty of Organization and Informatics Varazdin, 2022, pp. 349-356.
[45] T.-Y. Lin, M. Maire, S. Belongie, J. Hays, P. Perona, D. Ramanan, P. Dollár, C.L. Zitnick, Microsoft COCO: Common Objects in Context, in: D. Fleet, T. Pajdla, B. Schiele, T. Tuytelaars (Eds.) Computer Vision – ECCV 2014, Springer International Publishing, Cham, 2014, pp. 740-755.
[46] Tzutalin, LabelImg, Label Studio community, 2016.
[47] B. Dwyer, Nelson, J., Solawetz, J., Roboflow (Version 1.0) 2022. https://roboflow.com. (Accessed 2022.
[48] A. Bochkovskiy, C.-Y. Wang, H.-Y.M. Liao, Yolov4: Optimal speed and accuracy of object detection, arXiv preprint arXiv:2004.10934 (2020).
[49] R. Ribani, M. Marengoni, A survey of transfer learning for convolutional neural networks, 2019 32nd SIBGRAPI conference on graphics, patterns and images tutorials (SIBGRAPI-T), IEEE, 2019, pp. 47-57.
[50] M. Everingham, S.A. Eslami, L. Van Gool, C.K. Williams, J. Winn, A. Zisserman, The pascal visual object classes challenge: A retrospective, International journal of computer vision 111 (2015) 98-136.
[51] K.M. Ting, Precision and Recall, Encyclopedia of machine learning 781 (2010).
[52] M. Zhu, Recall, precision and average precision, Department of Statistics and Actuarial Science, University of Waterloo, Waterloo 2(30) (2004) 6.
[53] P. Henderson, V. Ferrari, End-to-end training of object class detectors for mean average precision, Computer Vision–ACCV 2016: 13th Asian Conference on Computer Vision, Taipei, Taiwan, November 20-24, 2016, Revised Selected Papers, Part V 13, Springer, 2017, pp. 198-213.
[54] P.E. Hart, How the Hough transform was invented [DSP History], IEEE Signal Processing Magazine 26(6) (2009) 18-22.
[55] R.O. Duda, P.E. Hart, Use of the Hough transformation to detect lines and curves in pictures, Communications of the ACM 15(1) (1972) 11-15.
[56] P. Soille, Erosion and Dilation, Morphological Image Analysis: Principles and Applications, Springer Berlin Heidelberg, Berlin, Heidelberg, 2004, pp. 63-103.
[57] Z. Xu, X. Baojie, W. Guoxin, Canny edge detection based on Open CV, 2017 13th IEEE international conference on electronic measurement & instruments (ICEMI), IEEE, 2017, pp. 53-56.
[58] G. Bradski, The OpenCV Library, Dr. Dobb's Journal of Software Tools (2000).
[59] F. Milletari, Hough voting strategies for segmentation, detection and tracking, Technische Universität München, 2018.
[60] H. Eibisch, A. Lohmuller, N. Kompel, R.F. Singer, Effect of solidification microstructure and Ca additions on creep strength of magnesium alloy AZ91 processed by Thixomolding, Int J Mater Res 99(1) (2008) 56-66.
[61] A. Akhtar, E. Teghtsoonian, Solid solution strengthening of magnesium single crystals—I alloying behaviour in basal slip, Acta Metallurgica 17(11) (1969) 1339-1349.
[62] H. Conrad, W.D. Robertson, Effect of temperature on the flow stress and strain-hardening coefficient of magnesium single crystals, JOM 9(4) (1957) 503-512.
[63] A. Chapuis, J.H. Driver, Temperature dependency of slip and twinning in plane strain compressed magnesium single crystals, Acta Materialia 59(5) (2011) 1986-1994.
[64] Q. Yu, J. Zhang, Y. Jiang, Direct observation of twinning–detwinning–retwinning on magnesium single crystal subjected to strain-controlled cyclic tension–compression in [0 0 0 1] direction, Philosophical Magazine Letters 91(12) (2011) 757-765.

[65] R.E. Reed-Hill, W.D. Robertson, Deformation of magnesium single crystals by nonbasal slip, JOM 9(4) (1957) 496-502.
[66] A. Akhtar, E. Teghtsoonian, Solid solution strengthening of magnesium single crystals—ii the effect of solute on the ease of prismatic slip, Acta Metallurgica 17(11) (1969) 1351-1356.
[67] T. Obara, H. Yoshinga, S. Morozumi, {ll-22}<-1-123> Slip system in magnesium, Acta Metallurgica 21(7) (1973) 845-853.
[68] Y. Xue, N. Takata, H. Li, M. Kobashi, L. Yuan, Critical resolved shear stress of activated slips measured by micropillar compression tests for single-crystals of Cr-based Laves phases, Materials Science and Engineering: A 806 (2021) 140861.
[69] W. Luo, C. Kirchlechner, J. Zavašnik, W. Lu, G. Dehm, F. Stein, Crystal structure and composition dependence of mechanical properties of single-crystalline NbCo2 Laves phase, Acta Materialia 184 (2020) 151-163.
[70] S. Huang, Q. Zhao, C. Lin, C. Wu, Y. Zhao, W. Jia, C. Mao, In-situ investigation of tensile behaviors of Ti–6Al alloy with extra low interstitial, Materials Science and Engineering: A 809 (2021) 140958.
[71] S. Joseph, I. Bantounas, T.C. Lindley, D. Dye, Slip transfer and deformation structures resulting from the low cycle fatigue of near-alpha titanium alloy Ti-6242Si, International Journal of Plasticity 100 (2018) 90-103.
[72] S. Amerioun, S.I. Simak, U. Häussermann, Laves-Phase Structural Changes in the System $CaAl_{2-x}Mg_x$, Inorganic Chemistry 42(5) (2003) 1467-1474.
[73] S. Schröders, S. Sandlöbes, C. Birke, M. Loeck, L. Peters, C. Tromas, S. Korte-Kerzel, Room temperature deformation in the Fe7Mo6 µ-Phase, International Journal of Plasticity 108 (2018) 125-143.
[74] M. Zubair, M. Felten, B. Hallstedt, M.V. Paredes, L. Abdellaoui, R.B. Villoro, M. Lipinska-Chwalek, N. Ayeb, H. Springer, J. Mayer, Laves phases in Mg-Al-Ca alloys and their effect on mechanical properties, Materials & Design 225 (2023) 111470.
[75] W. Luo, Z. Xie, P.-L. Sun, J.-L. Gibson, S. Korte-Kerzel, Plasticity of the Nb-rich µ-Co7Nb6 phase at room temperature and 600° C, Acta Materialia (2023) 118720.
[76] A. Foden, A. Previero, T.B. Britton, Advances in electron backscatter diffraction, arXiv preprint arXiv:1908.04860 (2019).